# A Test Statistic Estimation-based Approach for Establishing Self-interpretable CNN-based Binary Classifiers

Sourya Sengupta and Mark A. Anastasio *Fellow, IEEE*

*Abstract*—Interpretability is highly desired for deep neural network-based classifiers, especially when addressing high-stake decisions in medical imaging. Commonly used post-hoc interpretability methods have the limitation that they can produce plausible but different interpretations of a given model, leading to ambiguity about which one to choose. To address this problem, a novel decision-theory-inspired approach is investigated to establish a self-interpretable model, given a pre-trained deep binary black-box medical image classifier. This approach involves utilizing a self-interpretable encoder-decoder model in conjunction with a single-layer fully connected network with unity weights. The model is trained to estimate the test statistic of the given trained black-box deep binary classifier to maintain a similar accuracy. The decoder output image, referred to as an *equivalency map*, is an image that represents a transformed version of the to-be-classified image that, when processed by the fixed fully connected layer, produces the same test statistic value as the original classifier. The *equivalency map* provides a visualization of the transformed image features that directly contribute to the test statistic value and, moreover, permits quantification of their relative contributions. Unlike the traditional post-hoc interpretability methods, the proposed method is self-interpretable, quantitative. Detailed quantitative and qualitative analyses have been performed with three different medical image binary classification tasks.

## I. INTRODUCTION

Despite showing excellent potential for performing important tasks such as image classification and object detection, deep learning models are often criticized as being black-boxes that cannot be interpreted [1], [2]. However, such methods may not provide a unique interpretation of how the black-box models arrived at their decisions. This is because many convincing but different explanations or interpretations can be produced [3] and it is not always clear which interpretation is "correct" among them. This can clearly confound the goal of interpreting a black box model. There exist self-interpretable deep learning models, but many of them suffer from an interpretability-performance trade-off [3]–[5]. Hence, there is an urgent need for the development of alternative methods for achieving self-interpretability that can maintain the performance of a black-box classifier.

In this work, the following problem is addressed: *Given a trained deep binary black-box medical image classifier and the training images, find an alternative self-interpretable network that can deliver comparable classification accuracy.* To accomplish this, the original network is re-expressed in the form of an encoder-decoder model coupled with a single-layer fully connected network with unity weights. This model is trained in such a way that the output of the decoder, referred to as an *equivalency map*, represents a transformed version of the to-be-classified image whose element-wise sum approximates the same test statistic value as the original classifier. As such, the *equivalency map* provides a quantitative and novel means of understanding how the transformed image features contribute to the test statistic value.

Unlike traditional post-hoc interpretability methods, our approach is inspired from decision theory and it aims to establish a self-interpretable binary classifier. In decision theory, a classifier computes a scalar-valued test statistic from the input image, which is subsequently thresholded to make a decision. In our self-interpretable model, the *equivalency map* captures the transformation of the input image features to an image whose elementwise sum approximates the same test statistic value yielded by the original classifier. The proposed method has been rigorously evaluated through detailed quantitative and qualitative analyses of three different medical image binary classification tasks. Some distinctive characteristics of the proposed method are:

- It is based on a novel decision theory-inspired framework for developing self-interpretable models for medical image classification tasks.
- The proposed self-interpretable classifier achieved a classification accuracy that is on par with the original black-box classifier. This demonstrates the effectiveness of our self-interpretable model in achieving high-performance results while providing interpretability.

This work was supported in part by NIH Awards EB031772 (subproject 6366), EB031585 and CA238191. Research reported in this publication was supported by the National Institute Of Biomedical Imaging And Bioengineering of the National Institutes of Health under Award Number T32EB019944. The content is solely the responsibility of the authors and does not necessarily represent the official views of the National Institutes of Health.

Sourya Sengupta is with the Department of Electrical and Computer Engineering, University of Illinois Urbana–Champaign, Urbana, IL 61801 USA (e-mail: souryas2@illinois.edu).

Mark A. Anastasio is with the Department of Bioengineering, University of Illinois Urbana–Champaign, Urbana, IL 61801 USA (e-mail: maa@illinois.edu).

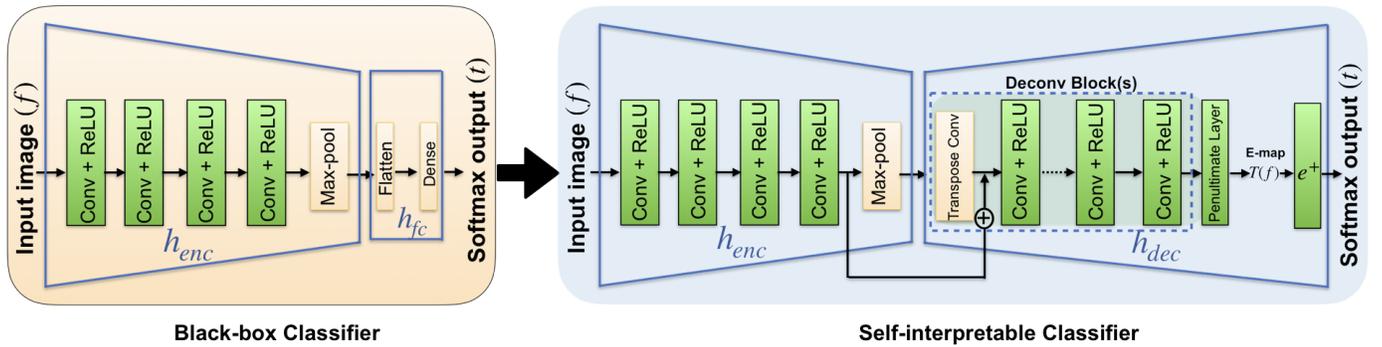

Fig. 1: The black-box classification network (left) and self-interpretable model involving an encoder-decoder network (right)

## II. BACKGROUND

### A. Post-hoc Interpretability Methods

Traditional post-hoc interpretability methods for black-box deep learning classifiers typically involve analyzing the model's output and its relationship to the input data. Popular such methods include gradient-based class activation maps (CAMs) [1]. These involve computing the gradient of the output with respect to the input features to identify which parts of the input are most important for a given prediction, which is typically visualized as a heatmap that highlights the important regions of the input. Some examples of these methods include the Saliency map [6], Guided Backprop [7], Gradient-weighted Class Activation Mapping (Grad-CAM) [8], Integrated Gradients [9], LIME [10] and the Layer-wise Relevance Propagation (LRP) [11]. However, these methods can produce different visualizations for the same black-box classifier [3]. Additionally, the interpretation of the heatmaps may not always be straightforward, and it may be difficult to determine which features or regions of the input are truly important for a given prediction.

### B. Self-interpretable Methods

Self-interpretable deep learning-based classifiers possess built-in interpretability components in the network architecture or training scheme, eliminating the need for traditional post-hoc methods [3]–[5]. Several models, including FRESH [12], SENN [13], Concept Bottleneck Models [14], ProtoPNet [15], and NAM [16], provide interpretations in different ways. For instance, FRESH focuses on interpretability for natural language processing tasks, while SENN and Concept Bottleneck Models generate interpretations in high-level spaces instead of raw pixel space. ProtoPNet provides interpretations in the pixel space, but with a focus on local patches that correspond to local areas of an image rather than global interpretation. NAM provides the same type of interpretations as SITE, but it combines neural networks with additive models to facilitate self-interpretation via component function. However, a drawback of many available self-interpretable models is that they may sacrifice classification performance [4], [17]. Chidester et al. [18] proposed a rotation equivariant CNN-based classifier that was found to learn more interpretable feature maps than those produced by a traditional CNN. However, this work only addressed the rotation-equivariance of feature maps of particular convolutional layers of the classifier.

## III. METHODOLOGY

From the perspective of decision theory, a binary classification of an image $f \in \mathbb{R}^N$ involves computation of a scalar-valued test statistic $t = h(f)$, where $h(f)$ is referred to as the discriminant function. For a linear classifier, the test statistic can be formulated as $t = h(f) = w^\dagger f + b$, where $w \in \mathbb{R}^N$ is called the decision template. Without loss of generality, we assume $b = 0$ in the discussion below. This mapping can alternatively be expressed as

$$t = w^\dagger f = e^\dagger [w \odot f], \quad (1)$$

where $e \in \mathbb{R}^N$ is a vector of all 1s and $\odot$ denotes the Hadamard product. This model is self-interpretable because $w \odot f$ can be readily visualized to understand the features employed to form the test statistic.

For a non-linear classifier, the test statistic $t$ can similarly be expressed as $t = h_{nl}(f)$, where the subscript $nl$ denotes that the discriminant function is non-linear. Inspired by Eq. (1), the test statistic for the non-linear classifier can be re-expressed as

$$t = h_{nl}(f) = e^\dagger T(f), \quad (2)$$

where $T : \mathbb{R}^N \to \mathbb{R}^N$ is a non-linear mapping that maps the input image $f$ into a transformed image $T(f)$. The test statistic value is computed by taking an element-wise summation of $T(f)$.

Consider that a deep neural network is employed to represent the discriminant function $h_{nl}(f)$. In this case, directly interpreting $h_{nl}(f)$ is known to be problematic. However, a key observation is that Eq. (2) provides a potentially interpretable alternative form of the black-box non-linear classifier. For a non-linear classifier, $T(f)$ can be thought of as a generalization of the quantity $w \odot f$ in Eq. (1). According to Eq. (2), $T(f)$ represents a transformed, or equivalent, version of the to-be-classified image that, when subject to an elementwise summation by a linear single layer neural network (SLNN) with unity weights, produces the test statistic value prescribed by the original discriminant function $h_{nl}(f)$. We therefore refer to $T(f)$ as an *equivalency map* (E-map). Because the formation of the test statistic via the SLNN is fully interpretable, the E-map provides a visualization of the transformed image features that contribute to the test statistic value and, moreover, permits quantification of their relative contributions. Below, the means by which the E-map can be computed is described.

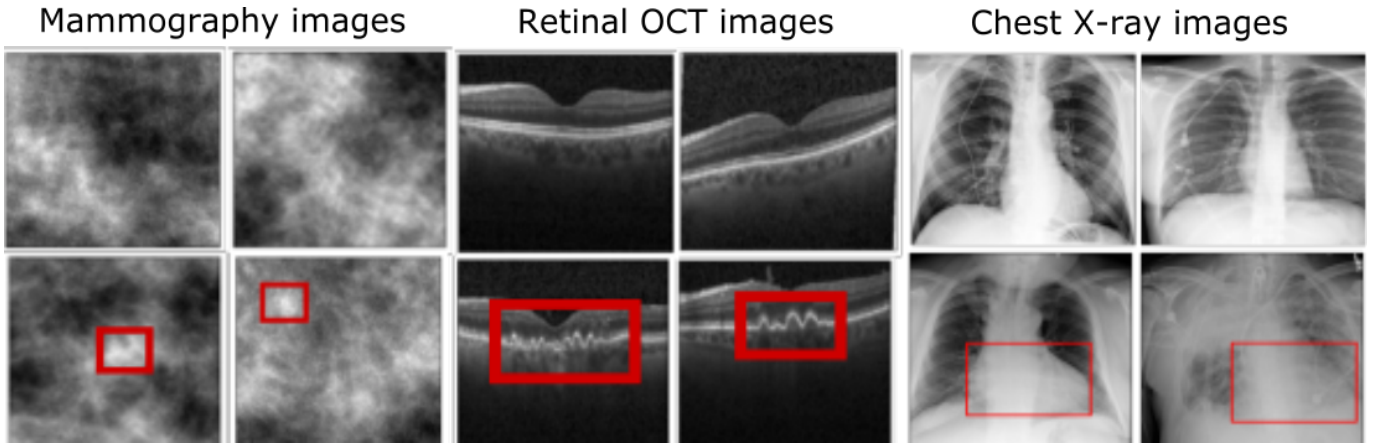

Fig. 2: Example image of all datasets. Top row: Normal images, bottom row: Abnormal images. The red bounding boxes indicate the regions of abnormality.

**Equivalency Map Computation** Consider that a non-linear discriminant function $h_{nl}(f)$ is represented as a composition of a feature-extracting encoder network ($h_{enc}$) and a fully connected network ($h_{fc}$):

$$t = h_{nl}(f) \equiv h_{fc}(h_{enc}(f)). \quad (3)$$

This configuration is referred to as the 'original' classifier, which is assumed to be trained and provided. As depicted in Fig. 1, the key contribution of this work is to establish an alternative configuration of the original classifier, henceforth termed as the self-interpretable network or interpretable encoder-decoder network, which can be interpreted via an E-map according to Eq. (2). To accomplish this, we approximate $T(f)$ in Eq. (2) by use of an encoder-decoder network, where the encoder is non-trainable and corresponds to $h_{enc}$ employed by the original classifier. Hence, only the decoder network is trainable, and the decoder output, the E-map ($T(f)$), can be approximated as

$$T(f) \approx h_{dec}^{\theta^*}(h_{enc}(f)), \quad (4)$$

where $h_{dec}^{\theta^*}(\cdot)$ represents the decoder network parameterized with weights $\theta^*$. The decoder parameters are estimated in a way so that the self-interpretable network learns to estimate the test statistic $t$ of the original classifier. Specifically, the decoder parameters are estimated in such a way that $e^\dagger h_{dec}^{\theta^*}(h_{enc}(f)) \approx h_{nl}(f) = t$.

Hence, the decoder network is trained by (approximately) solving the following optimization problem:

$$\theta^* = \underset{\theta}{\operatorname{argmin}} \underset{f \sim D}{E} \left\{ L(h_{nl}(f), e^\dagger h_{dec}^\theta(h_{enc}(f))) \right\}. \quad (5)$$

Here, $D$ denotes the distribution of the training images $f$ and $L$ denotes the loss function, which corresponds to mean squared error (MSE) in the studies below. The equivalency map computation steps are shown in the summary below.

---

**Summary: Procedure for Computing Equivalency Maps**

**Input:** Non-linear black-box classifier $h_{nl}(f)$, training data $f$

Let $h_{enc}$ be the encoder network in the original classifier
Let $h_{fc}$ be the fully connected network in the original classifier
Compute: $t \leftarrow h_{nl}(f) \equiv h_{fc}(h_{enc}(f))$  ▷ Original classifier
Initialize trainable decoder network parameters $\theta$
Train the decoder network $h_{dec}^\theta$ by solving:
$\theta^* \leftarrow \underset{\theta}{\operatorname{argmin}} \underset{f \sim D}{E} \left\{ L(h_{nl}(f), e^\dagger h_{dec}^\theta(h_{enc}(f))) \right\}$
Compute the self-interpretable network's E-map as:
$T(f) \approx h_{dec}^{\theta^*}(h_{enc}(f))$
**Output:** Self-interpretable network and its E-map $T(f)$

---

## IV. EXPERIMENTS

Three different binary classification tasks were considered to evaluate and investigate the classification performance of the self-interpretable networks in terms of accuracy. Quantitative analyses were also performed to understand the pixel intensity distribution of the E-maps and the overlap between the disease area and contributing pixel locations. This allowed for a deeper understanding of the network's decision-making process and which features of the E-map were most relevant in determining the output class.

### A. Classification Tasks

Three different binary classification tasks were considered in our studies. Sample images from all the datasets are shown in Fig. 2, where the red bounding boxes are annotations that indicate the specific region where the abnormality is present.

**Drusen detection task using retinal OCT images:** A drusen detection task was performed using optical coherence tomography (OCT) images of the human retina of size 256 x 256 [19]. Drusen is characterized as an accumulation of extra-cellular materials between the retinal pigment epithelium (RPE) layer and the Bruch's membrane layer of the human retina and can be well observed using retinal OCT images.

**Tumor detection task using simulated mammography images:** A stylized tumor detection task was explored using a simulated digital mammography dataset. The doubiso clustered lumpy backgrounds (CLB) were used as background images [20]. The to-be-detected tumor was generated as a 2D symmetric Gaussian function and was inserted [21] into the background in one of the 9 discrete locations shown in Fig. 3. The images were of size 128 X 128.

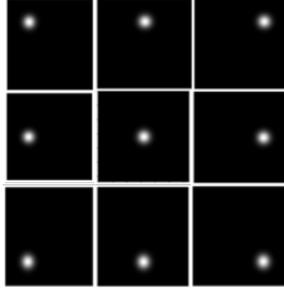

Fig. 3: The simulated tumor and different locations where it was inserted in the simulated CLB images.

**Cardiomegaly detection task using chest X-ray images:** A cardiomegaly detection task was performed using chest X-ray images of size 1024 X 1024 images. Cardiomegaly refers to the enlargement of the heart, which is a biomarker for heart diseases. The images were taken from a publicly available NIH database [22]. The image labels were created using text mining from radiological reports generated by clinicians.

### B. Training Details

*1) Black-box classifiers:* For the black-box classifier, two different CNN configurations were used in our experiments for all three tasks. The first classifier (baseline) consisted of 4 convolutional blocks (convolution + non-linear activation) followed by a max-pool layer and a fully connected dense layer. The VGG16 network [23] was used as another black-box deep network.

*2) Self-interpretable networks:* The proposed self-interpretable network has an encoder-decoder style architecture. The feature extraction component of the corresponding black-box network is employed as the pre-trained encoder of the self-interpretable network. The decoder comprises several components, including convolutional layers, transpose convolutional layer, and skip connection, grouped together as a unit referred to as a 'Deconv block'. The Decoder can have single or multiple such Deconv blocks depending upon the design. The number of Deconv blocks is the same as the number of maxpool layers in the encoder to keep the spatial dimension of the decoder output similar to the input image. The details of the decoder architecture of the network are summarized here and also shown in the Table I

**Deconv block**:
1) The Deconv block starts with a deconvolutional or transpose convolutional layer that has 128 filters and a kernel size of (2, 2). It employs strides of 2 to up-sample the input.
2) The output of the previous layer is then concatenated with the corresponding similar-size output from the encoder layer with a skip connection.
3) Next, a sequence of multiple consecutive convolutional layers is applied. Each layer has 128 filters and a 5x5 kernel size. The ReLU activation function is used for each layer.

**Penultimate layer:** After the deconv block(s), a final convolutional layer is applied. This layer has 1 filter and a 5x5 kernel size. The activation function used is the non-linear activation function ReLU.

Finally, a dense layer is used to take the element-wise summation of the output of the decoder or E-map.

TABLE I: Decoder architecture details

|  | Type | Number of Feature Maps | Kernel Size | Stride | Activation |
|---|---|---|---|---|---|
| Deconv Block | Transposed Convolution | 128 | (2,2) | 2 | ReLU |
|  | Skip Connection | - | - | - | - |
|  | Convolutional Layer | 128, 128, 128 | (5,5) | 1 | ReLU |
| Penultimate Layer | Convolutional Layer | 1 | (5,5) |  | ReLU |

*3) Training:* For the tumor detection and drusen detection tasks, the training, validation, and testing sets comprised 19000, 1000, and 1000 images in each class, respectively. For the drusen detection task, the training, validation and testing sets comprised 6000, 1000, and 1000 images in each class respectively. For the cardiomegaly detection task, 2000, 200 and 200 images were used for the training, validation, and testing respectively. Binary cross-entropy was used as the loss function for the black-box classifiers and mean squared error (MSE) was used as the loss function for the self-interpretable network. The Adam optimizer [24] with a learning rate of 3e-5 was used to train all the models. A stopping rule was designed to stop the training if the validation loss did not decrease for consecutive five epochs.

### C. Performance of the self-interpretable network

For all the tasks, the classification accuracy of the black-box classifier and the self-interpretable classifier, along with test statistic estimation errors were computed. The baseline CNN achieved test statistic estimation errors of 0.001, 0.0003, and 0.003 for the mammography, OCT, and chest X-ray datasets, respectively. The VGG16 model had estimation errors of 0.003, 0.0005, and 0.003 for the mammography, OCT, and chest X-ray datasets, respectively. It was observed that the accuracy achieved by the self-interpretable network was similar to the original classifier for all the cases. Table II and III contain the classification accuracies for all three tasks for both baseline CNN (4-layer) and VGG16. The ROC curves for the tumor detection and cardiomegaly detection tasks are shown in Fig. 4 and Fig. 5. The AUC value for the drusen detection task was nearly 1. The near-ideal ROC curves for that task are not displayed here. It was observed that the self-interpretable classifiers yielded ROC curves that closely approximated those yielded by the black-box classifiers. The AUC values were also comparable between black-box and self-interpretable classifiers.

TABLE II: Classification accuracy (%) of the baseline CNN classifier and the corresponding self-interpretable network for 3 different tasks. Both networks achieved similar classification accuracy.

| Dataset | Classification Accuracy of the Baseline Black-box Classifier | Classification Accuracy of the Associated Self-interpretable Network |
|---|---|---|
| Mammography | 77.8 | 77.8 |
| Retinal OCT | 99.1 | 99.1 |
| Chest X-ray | 83.33 | 83.0 |

TABLE III: Classification accuracy (%) of the VGG16 classifier and the corresponding self-interpretable network. Both networks achieved similar classification accuracy.

| Dataset | Classification Accuracy, Sen, Spec of the VGG16 Black-box Classifier | Classification Accuracy, Sen, Spec of the Associated Self-interpretable Network |
|---|---|---|
| Mammography | 79.8 | 79.8 |
| Retinal OCT | 99.5 | 99.5 |
| Chest X-ray | 81.2 | 81.0 |

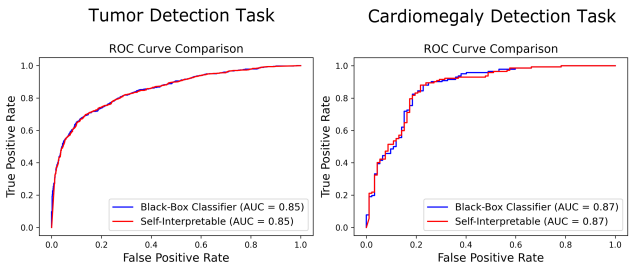

Fig. 4: ROC curves corresponding to the baseline CNN classifier (blue) and the corresponding self-interpretable classifier (red) for different tasks. For both tasks, the ROC curves corresponding to the self-interpretable classifier closely approximated those corresponding to the black-box classifier.

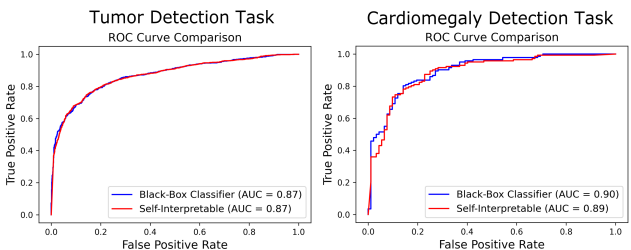

Fig. 5: ROC curves corresponding to the VGG16 classifier (blue) and the corresponding self-interpretable classifier (red) for different tasks. For both tasks, the ROC curves corresponding to the self-interpretable classifier closely approximated those corresponding to the black-box classifier.

### D. Visualizing Equivalency Maps

Figure 6 shows an abnormal mammography image and examples of corresponding heatmaps generated by different state-of-the-art post-hoc interpretability methods for the trained baseline CNN classifier for the normal vs tumor mammography classification task.

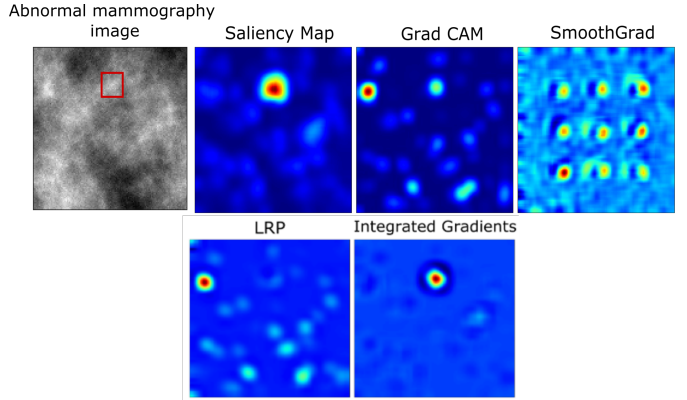

Fig. 6: Heatmap interpretations generated by different post-hoc interpretability methods for the baseline CNN classifier. The results show how different methods can yield multiple plausible but different visualizations.

The Saliency map [6], Integrated Gradients (IG) [9], Guided Backprop [7], Grad-CAM [8], LRP [11], Smoothgrad [25] were used to interpret the classifier. It was observed that different methods could yield multiple plausible but different visualizations when an abnormal mammography image was considered, which can confound model interpretation.

Figure 7 shows E-maps generated by the self-interpretable network, whose encoder was fixed and specified by the feature extraction layer weights of the baseline CNN classifier. The E-maps were overlayed with the original images for abnormal classes for all the datasets. The E-maps tend to reveal relevant regions where the abnormality is present in the abnormal images. It was also observed that, for the abnormal images, the E-map tended to have positive values (bright pixels) at the locations of abnormal features. These pixels contributed significantly to the test statistic, yielding relatively large test statistic values that resulted in the classification of the images as abnormal. On the other hand, the images from normal class did not show specific patterns and yielded lower test statistics values, as shown in the Appendix A. E-maps for the VGG16 network are shown in the Appendix B.

### E. Performance with Different Number of Layers in Baseline Black-box Classifier

In this study, the performances of both the baseline black-box classifier and the corresponding self-interpretable classifier were explored with consideration of various number of convolutional layers in the black-box classifier when the decoder of the self-interpretable classifier was fixed. The baseline black-box classifier utilized multiple convolutional layers, followed by a max-pooling layer, and concluded with a final dense layer. For the experiments, performances with a total of

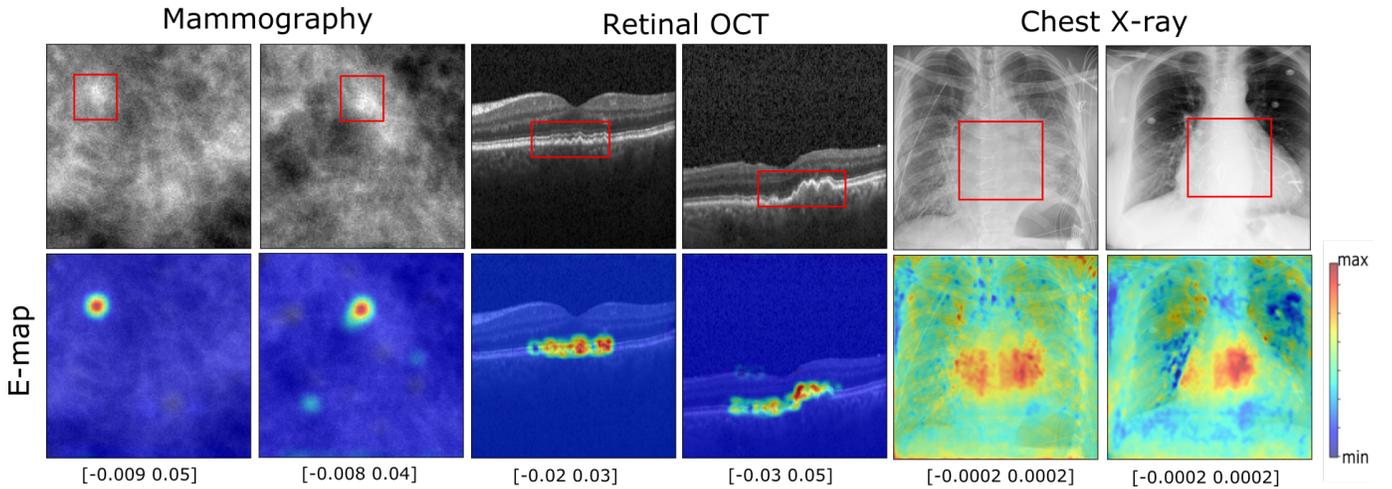

Fig. 7: Top row: Sample images for abnormal classes from the datasets. Bottom row: The corresponding E-maps overlayed on the original image. The E-maps tend to show regions where an abnormality is present. The red bounding boxes show the region of the abnormality. The pixel intensity value range of the E-map is shown below for each E-map. The colorbar is shown on the right and applies to all E-maps.

1, 3, 4, and 6 convolutional layers were analyzed. In the corresponding self-interpretable classifier, the encoder corresponded to the black-box model's feature extraction network, excluding the final dense layer. The decoder consisted of a single Deconv block, which consisted of one transpose convolutional layer, a skip connection with a corresponding convolutional layer from the encoder and multiple subsequent convolutional layers. The Deconv block was followed by a penultimate convolutional layer, and a final dense layer to sum the decoder output or E-map. Table IV presents the classification accuracy results for different number of convolutional layers in the baseline CNN classifier. Notably, the self-interpretable classifier consistently achieved the same accuracy for different baseline black-box classifiers with different number of layers.

TABLE IV: Classification accuracy (%) of the baseline CNN classifier and self-interpretable network with consideration of different numbers of convolutional layers in the baseline CNN.

|  | Mammography | | OCT | | X-ray | |
| --- | --- | --- | --- | --- | --- | --- |
|  | Black-box | Self-interpretable | Black-box | Self-interpretable | Black-box | Self-interpretable |
| 1 Conv Layer | 70 | 70 | 91 | 91 | 71.8 | 71.7 |
| 3 Conv Layers | 76.2 | 76.2 | 94.6 | 94.6 | 81.6 | 81.6 |
| 4 Conv Layers | 77.8 | 77.8 | 99.1 | 99.1 | 83.33 | 83 |
| 6 Conv Layers | 77.9 | 77.9 | 99.5 | 99.5 | 79.1 | 79 |

### F. Performance for Different Number of Decoder Layers in Self-interpretable Classifier

In this study, the influence of the decoder architecture was investigated when the corresponding black-box classifier and the encoder of the self-interpretable classifier were fixed. The decoder architecture in the self-interpretable classifier was investigated for both the baseline black-box classifier and the VGG16 classifiers. The cardiomegaly detection task involving the chest X-ray dataset was employed in this study. The primary focus was to analyze the performance with different number of convolutional layers within each Deconv block of the decoder in the self-interpretable classifier. Experiments were conducted with varying configurations that included 0, 2, 3, 5, and 7 convolutional layers within each Deconv block of the decoder. The baseline black-box classifier contained 4 convolutional layers, followed by a max-pooling layer, and concluded with a final dense layer. In the corresponding self-interpretable classifier, the encoder consisted of the architecture employed by the black-box model's feature extraction network, excluding the final dense layer. The decoder, on the other hand, consisted of Deconv blocks, with each block containing one transpose convolutional layer, a skip connection that paired with a corresponding convolutional layer from the encoder, and multiple additional convolutional layers. The Deconv block was followed by a penultimate convolutional layer and a final dense layer.

TABLE V: Classification performance (%) with different number of convolutional layers in the Deconv block of the self-interpretable classifier's decoder

| Task: Cardiomegaly Detection | 0 | 2 | 3 | 5 | 7 |
| --- | --- | --- | --- | --- | --- |
| Accuracy | 61 | 64.2 | 79.3 | 83.0 | 83.1 |

The total number of Deconv blocks (or total transpose convolutions in the decoder) was determined by the total number of max-pool layers in the encoder to match the spatial dimension of the E-map with the original input image. Table V shows that the performance of the self-interpretable classifier did not improve significantly after 5 convolutional layers in the Deconv block. For this task, the corresponding baseline black-box classifier achieved an accuracy of 83.33%. Similar experiments were also performed when the black-box classifier was VGG16 and the results are provided in Appendix C.2 In that case, a similar result was observed in which the classification performance of the corresponding self-interpretable classifier was improved with an increasing number of convolutional layers. However, beyond 5 convolutional layers in each Deconv block, the performance did not exhibit significant improvement.

### G. Examination of False Positive (FP) and False Negative (FN) Cases

In the context of medical imaging, false positive (FP) and false negative (FN) cases hold considerable significance. As depicted in Figure 8, images and corresponding E-maps from the mammography and chest X-ray dataset are provided to illustrate FP and FN cases. Upon examination of the E-maps, it becomes evident that FN cases fail to accurately locate abnormal regions within the images. In contrast, FP cases exhibit E-maps that bear a closer resemblance to E-maps of the true positive class. Similar experiments were also performed for the case where the black-box classifier was VGG16. The results are shown in Appendix C.3 and the findings were qualitatively similar to those described above.

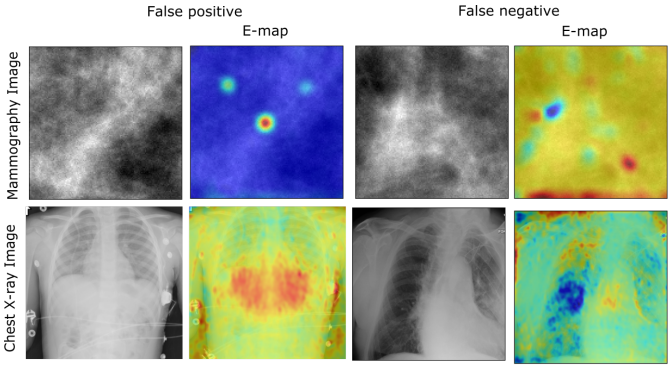

Fig. 8: False Positive (FP) and False Negative (FN) Analysis. FP cases look similar to positive class E-map, FN cases look similar to negative class E-map

### H. Stability Analysis of Equivalency Maps

In this study, the stability of the E-map was assessed for a given architecture of the self-interpretable network across different random weight initializations, considering both normal and uniform distributions. The stability of deep learning model interpretation is a critical factor in ensuring the trustworthiness and reproducibility of the results. We defined stability as the degree to which the same interpretation can be obtained from multiple runs with different random weight initializations for a given self-interpretable architecture. When a self-interpretable model produces similar interpretations across different runs, it is considered to be stable. To assess this, three binary classification tasks were considered with three different random weight initializations: 'glorot uniform' [26], 'random normal' and 'random uniform'. The self-interpretable models were trained for each of these conditions and E-maps were computed. While evaluating quantitatively, for all the tasks, the structural similarity index (SSIM) [27] values were computed. For the mammography and OCT cases, all three restarts showed high similarity with an SSIM score of 0.98 in both cases, indicating a high degree of stability in the interpretation of the model. For the cardiomegaly detection task that involved the chest X-ray dataset, the glorot uniform and random uniform weight initializations produced E-maps with a high similarity of 0.98 SSIM. The studies that employed initializations based on the random normal and glorot uniform distributions also achieved fairly good similarity of 0.85 between the two E-maps. Figure 9 shows some visualizations of the results. Qualitatively similar results were found when the pre-trained feature extraction network of the VGG16 classifier was used as the encoder of the self-interpretable network. Those results are presented in Appendix C.4.

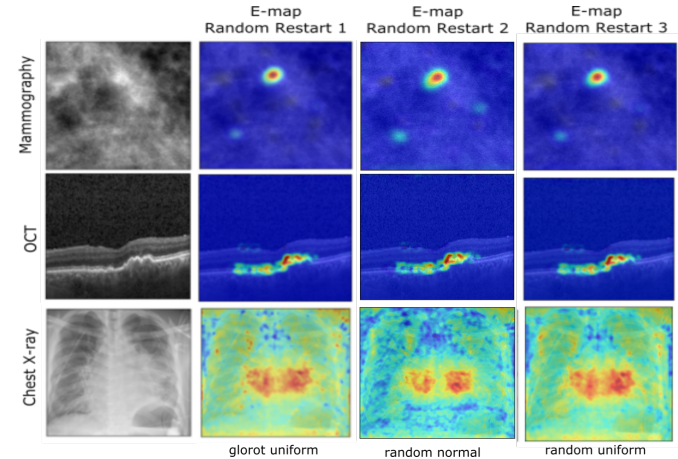

Fig. 9: From left in each row: input image and E-maps for 3 different random weight initializations of the network. Different random weight initializations produced similar-looking E-maps. This study shows the stability of the E-maps for different random restarts of the network.

### I. Pixel Intensity Distribution Analysis

In this study, an in-depth analysis of pixel intensity distributions in the E-maps was conducted to reveal variations between the E-maps of normal and abnormal class images. This analysis offers insights into the distinct contributions of different elements towards positive or negative decisions. The test statistic values for true positive cases were larger than true negative cases. As an elementwise sum of an E-map yields the test statistic, any positive element of an E-map contributes to classifying the image as abnormal. The negative elements act in a reverse way by minimizing the test statistic to predict the image as a normal case. In this study, this pixel intensity distribution analysis can reveal insights about how the pixel intensity distribution varies between the E-maps of normal and abnormal images and how the different elements contribute towards a decision in a positive or negative manner. In Fig. 10, the histogram is plotted for positively contributing pixels of the E-maps of normal and abnormal images of different tasks. It can be seen that there is a significant difference in positively contributing element values for abnormal cases compared to the normal images. Results corresponding to VGG16 can be found in Appendix C.5.

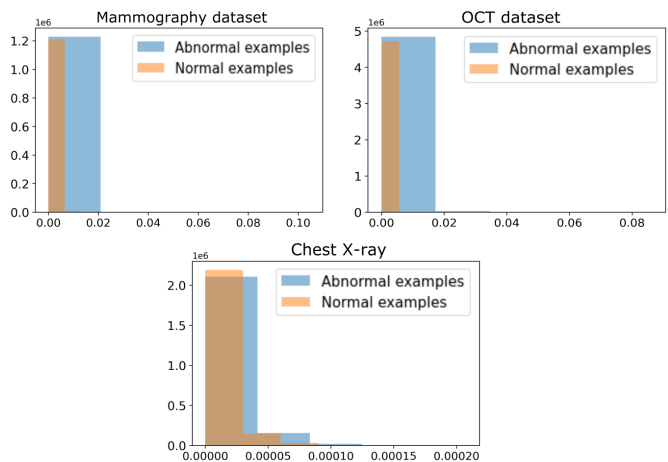

Fig. 10: The histograms of normal vs abnormal cases for each task. It can be seen there is a significant difference in positively contributing elements for abnormal compared to normal cases.

### J. Quantitatively Evaluating Interpretability : E-map Contributions from Abnormality Regions

In this study, interpretability was quantitatively evaluated by examining the contributions of E-map pixels to the test statistic value. The overlap between the regions of the top contributing E-map pixels and the actual abnormality regions was computed to assess the spatial correspondence of localization with relevant regions in the abnormal images. This quantitative approach allowed us to precisely measure the relation between contributing pixels and the test statistic value.

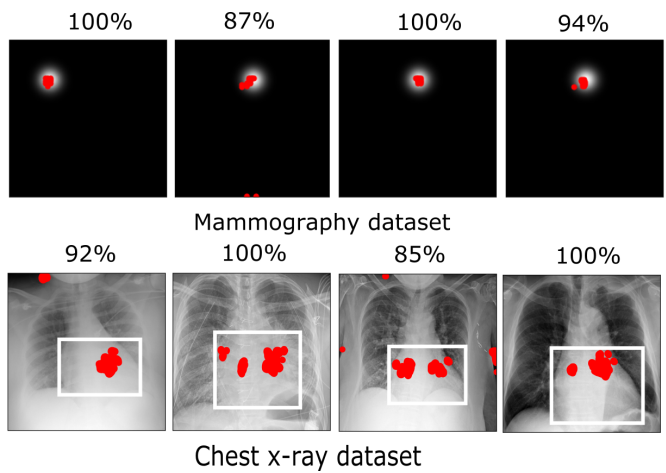

Fig. 11: Examples of overlap between the most contributing 1% elements of an E-map (in red) and the abnormal region. The percentage overlap is written above each image. In most cases, the E-map achieved a high overlap percentage.

The mammography dataset was simulated and hence the specific tumor regions were known. The NIH chest X-ray dataset had bounding box annotations for the cardiomegaly class. For these two datasets, the percentage overlap between the abnormal region and contributing pixels in each test set image was computed. This quantitative study shows how many top contributing pixels of an E-map overlap with the actual abnormality region. As our method is quantitative, the overlap between the disease region and contributing pixel toward test statistics can be quantitatively determined. Figure 11 reveals how the top 1% contributing pixels (red) overlap with the abnormal locations for mammography and chest X-ray datasets respectively for the baseline CNN network. The percentage overlap is written above each image. Results corresponding to the VGG16 network can be found in the Appendix C.6. The results for that also showed high percentages of overlap, similar to the results shown here.

A similar analysis was performed to compare our method with some commonly used post-hoc interpretability methods (Saliency map [6], Integrated Gradients (IG) [9], Guided Backprop [7], Grad-CAM [8], LRP [11]) of the corresponding black-box network in terms of quantitative performance of percentage overlap with abnormal regions. It is important to note here that these post-hoc methods are designed to explain an existing black-box classifier, whereas our method aims to establish a self-interpretable model that can also achieve similar classification accuracy with a black-box classifier. As the chest X-ray dataset (associated with the cardiomegaly detection task) has radiologists' annotations, this dataset was used to compare the methods in a quantitative manner. Figure 12 shows some examples of percentage overlap between the top 1% contributing pixels of the post-hoc interpretability heatmaps for the black-box classifier and the abnormal region of the original image. A similar analysis was done for the E-maps. It was observed that in most cases percentage overlap was higher in the E-map than in most of the post-hoc interpretability methods. It should also be noted how interpretations can vary for a single black-box classifier, which can be a potential issue in deciding which method to rely upon. Table IV shows a population level analysis of average percentage overlap between the top 1% pixels of the E-maps of our encoder-decoder based models and the post-hoc interpretability methods for corresponding black-box classifier with clinically annotated regions in the original abnormal images over all 100 test images. The superiority of our method can be shown from the values in the Table VI

TABLE VI: Average percentage overlap with the top 1% pixels of interpretation maps and clinically annotated regions in the original abnormal images- population-level analysis over all test images

|  | E-map | Guided Backprop | IG | LRP | Saliency Maps | Grad Cam |
|---|---|---|---|---|---|---|
| Percentage Overlap | 89% | 71% | 84% | 83% | 30% | 12% |

### K. Effect of Direct Training of self-interpretable network

In our training framework, the self-interpretable network employed a pre-trained encoder and it was trained with the objective of closely estimating the test statistic values produced by the original black-box classifier. An alternative training approach is to directly train the proposed encoder-decoder based network from scratch by the use of the original image labels 0,1 and a classification loss. In this case, the model involves random initialization of both the encoder and decoder, as a fixed pre-trained encoder is not utilized. While direct training of the proposed model may yield a similar level of interpretability, a scenario was identified where it degrades classification accuracy. For the cardiomegaly detection task, Table V shows how direct training of the network with a classification loss and 0,1 labels can result in degraded performance compared to the original black-box model. On the other hand, the proposed training scheme achieved a similar level of classification accuracy compared to the traditional black-box network. A possible reason for this behavior is that the effect of pre-training provides a better initialization of the self-interpretable network.

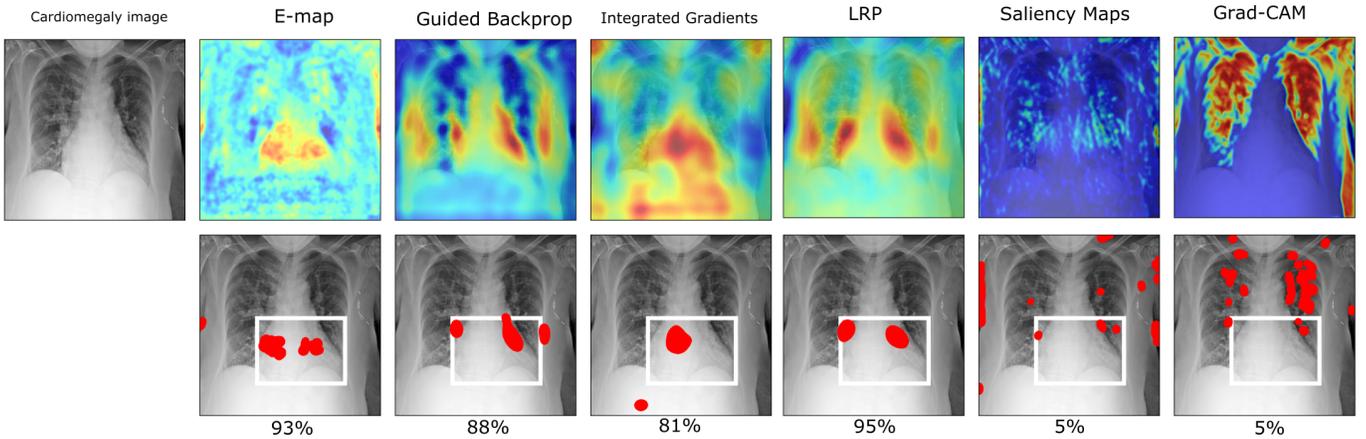

Fig. 12: Top row: E-map and heatmaps from different post-hoc interpretability methods. Bottom row: Percentage overlap with the top 1% pixels in the E-map(shown in red) and clinically annotated regions in the original abnormal image. The numbers signify the percentage overlap. It can be seen how different post-hoc interpretability methods can produce different-looking heatmaps. The percentage overlap with the top 1% pixels (red) and clinically annotated regions in the original abnormal image is higher than other post-hoc methods for the black-box network.

TABLE VII: Accuracy (%) of direct training of the self-interpretable network. This study shows how directly training the self-interpretable network with 0-1 labels and without pre-training can affect the classification performance.

| Task | Classification Accuracy of the VGG16 Black-box Classifier | Classification Accuracy of the Associated Self-interpretable Network | Classification Accuracy of Direct Training of the Associated Self-interpretable Network |
|---|---|---|---|
| Cardiomegaly Detection | 83.33 | 83 | 73.2 |

L. *Comparative Analysis with Competing Self-interpretable Methods*

Several self-interpretable methods, each employing novel self-interpretability strategies, were selected for comparison with our proposed method. Li et al. [28] introduced the self-interpretability method PrototypeDL. This method incorporates an autoencoder and a specialized prototype layer, enabling the network to provide explanations for its predictions. Through a multi-objective training approach, it learns prototypes that offer insights into the reasoning behind each prediction. ProtoPNet [15] is another state-of-the-art self-interpretable model in image classification, developed in a subsequent study. In that approach, the network comes to a decision by finding prototypical parts of an image, which is the key interpretability component. Recently, the Self-Interpretable Model with Transformation Equivariant Interpretation (SITE) [4] was introduced as a self-interpretable model that employs transformation equivariant regularization to learn robust interpretations. The model captures valid interpretations that are invariant to geometric transformations. As the interpretability formulations of these different methods differ fundamentally from our approaches, a direct comparison of interpretability between our method and these methods is challenging. However, as all the methods address classification problems, the classification performances can be compared. Table VIII shows that the alternate self-interpretable methods achieved lower classification performance compared to our model for the cardiomegaly detection task. This is consistent with the previously reported accuracy-interpretability trade-off with self-interpretable methods [17]. Notably, here, our self-interpretable model used a feature extraction component of the pre-trained VGG16 classifier as the encoder of the self-interpretable network.

TABLE VIII: Classification accuracy (%) and training time (in hours) of the black-box classifier, corresponding self-interpretable network, PrototypeDL, ProtoPNet and SITE. The results show that the performance of the other self-interpretable methods is lower than the proposed method which maintains the same level of accuracy as the original black-box classifier.

| Task: Cardiomegaly Detection | Classification Accuracy of the Black-box Classifier | Classification Accuracy of the Self-interpretable Network | PrototypeDL | ProtoPNet | SITE |
|---|---|---|---|---|---|
| Accuracy | 81.23 | 81.0 | 73.5 | 75.3 | 71 |
| Training Time | - | 5 | 3.5 | 6 | 4.1 |

It is to be noted that the learning rates of the competing methods were fine-tuned, but other parameters and hyper-parameters for all of those methods were adopted as specified in their respective research papers. All the models were trained using 2 GeForce GTX 1080 Ti GPUs with 12GB RAM. A stopping rule was designed to stop the training if the validation loss did not decrease for five consecutive epochs. The training times for the different approaches are also listed in Table VIII.

## V. Discussion

A novel decision-theory-inspired method was established to provide an alternative means of self-interpretability for binary medical image classification. The proposed method involves training an encoder-decoder-based model followed by a non-trainable fully connected layer with fixed unity weights. This network employed a pre-trained encoder from a black-box classifier and the model was trained using an estimation task to estimate the test statistic to maintain the performance of a given trained black-box deep binary classifier. By construction, the element-wise summation of the decoder output of the interpretable network (E-map) represents the test statistic value.

Self-interpretability of our method is derived from the direct interpretation of the test statistic formation from the E-map. This means that each element in the E-map contributes directly to the test statistic, thereby providing valuable insights into the underlying decision making of the network. The E-map is an image and may look qualitatively similar to CAMs visualizations in some situations. However, our method possesses significant differences from post-hoc interpretability methods in terms of formulation. It is important to note that the proposed method does not seek to interpret a black-box network. Rather, it seeks to establish an alternative self-interpretable network that closely mimics the classification performance of a given black-box model. This is a fundamental hallmark of the method.

It should also be noted that there is no available theoretical guarantee that the E-map $T(f)$ will always provide an interpretable visualization of the spatial signatures (features) in the original image $f$ that are utilized by the classifier. It is possible that there could be some applications in which the E-map does not accurately localize features in the original image. Though, in our studies conducted to-date, we have not observed this. Instead, it was found that the E-map $T(f)$ generally reveals regions in $f$ where the abnormality is present, offering a deeper understanding of the decision-making of the network.

One possible way to extend the proposed approach for use with multi-class problems may be to train multiple self-interpretable networks. If we have a total of $N+1$ classes in a task, a total of $N$ self-interpretable networks could be trained. In this way, each self-interpretable classifier would estimate the test statistic for class $n$ where $n = 1, ..., N$. The predicted class for a particular sample can be determined by finding the maximum value among these estimated test statistics.

## Appendix
### A. Normal Class E-maps for Network with baseline CNN Encoder

In this subsection, the E-maps of the normal class are shown when the encoder of the self-interpretable network used the feature extraction component of the pre-trained baseline black-box classifier. The E-maps for the normal class are shown in Fig. 13. Here the pre-trained baseline CNN was used as the encoder of the self-interpretable network. It can be seen that the E-maps for normal class images, do not show specific patterns, unlike the disease class.

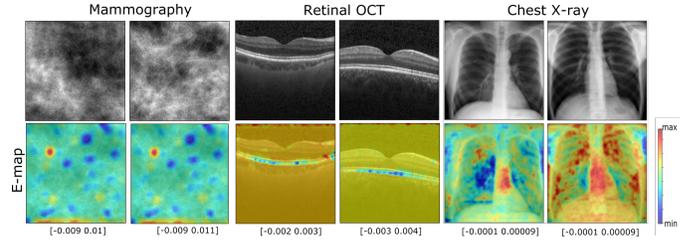

Fig. 13: Top row: Sample images for normal classes from the datasets. Bottom row: The corresponding E-maps overlayed on the original image.

### B. Presence of Multiple Anomalies

Previous results showed cases where a single anomaly was present in one abnormal image. In this study, a more complicated dataset was chosen with two abnormal regions present in a single image. The new dataset had the task of identifying tumors, the doubiso clustered lumpy backgrounds (CLB) were used as background images [20]. The to-be-detected tumor was generated as a 2D symmetric Gaussian function and was inserted [21] into the background.

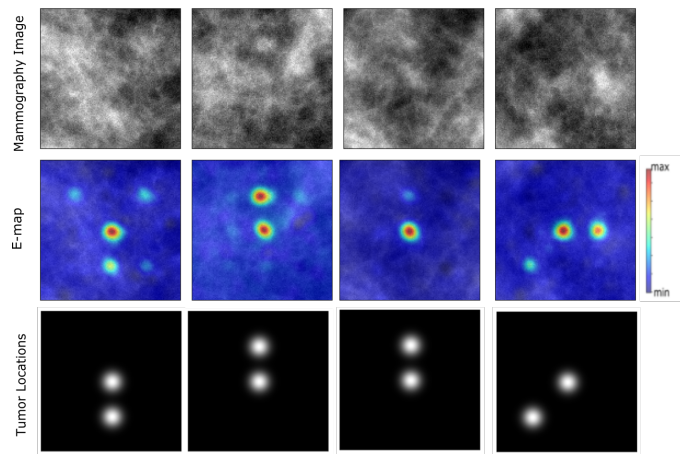

Fig. 14: Top row: Sample images for abnormal classes from the mammography dataset. Middle row: The corresponding E-maps overlayed on the original image. Bottom row: The tumor locations.

Notably, each abnormal class image now comprises two distinct tumors: one that varied in position (located at any of the nine predefined positions, similar to the previous tumor detection task dataset), and another, that was consistently positioned in the center. First, the baseline black-box classifier was trained to classify between tumor and normal classes. Subsequently, the corresponding self-interpretable classifier was trained to estimate the test statistics given by the original black-box classifier. The self-interpretable classifier was able to achieve comparable accuracy as the baseline black-box classifier. Figure 14 shows that the E-map mainly focused on the location of the middle fixed tumor in most of the cases. This is an intuitive result because that tumor location remained fixed, the model found it useful to exploit that information as opposed to trying to detect the randomly located tumor.

### C. Results of VGG16 Network

Results of the self-interpretable network associated with baseline CNN are shown in Sec. IV. Similarly, the results of the self-interpretable network corresponding to the VGG16 classifier are shown in this section. The black-box classifier was VGG16 and the pre-trained VGG16 feature extraction network was employed as the encoder of the self-interpretable network.

*1) Visualizing Equivalency Maps:* Figures 15, 16, and 17 present the E-maps for the mammography, OCT, and chest X-ray datasets, respectively, with each figure showcasing two images from the normal and abnormal classes. The feature extraction component of the trained VGG16 was used as the encoder. Consistent with the findings of the CNN architecture, the E-maps of the abnormal class highlight relevant regions where the abnormality is present. This finding is similar to the results discussed in Sec. IV.D. However, the normal class E-maps do not show any specific pattern.

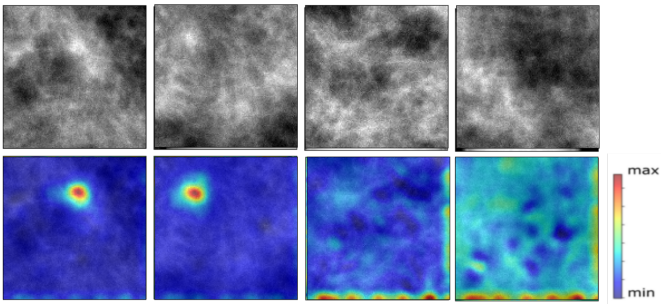

Fig. 15: Mammography E-maps (Two abnormal and two normals) of the self-interpretable network. Top row: the original images, Bottom row: E-maps. The abnormal class E-maps tend to show regions where an abnormality is present.

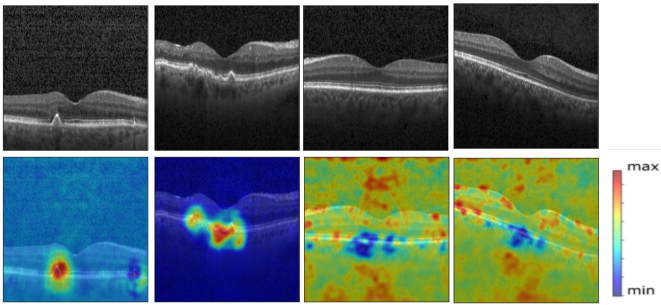

Fig. 16: OCT E-maps (Two abnormal and two normals) of the self-interpretable network. Top row: The original images; Bottom row: E-maps. The abnormal class E-maps tend to show regions where an abnormality is present.

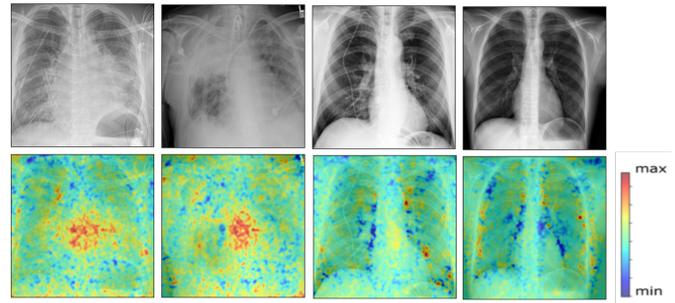

Fig. 17: Chest X-ray E-maps (Two abnormal and two normals) of the self-interpretable network. Top row: The original images, Bottom row: E-maps. The abnormal class E-maps tend to show regions where an abnormality is present.

*2) Performance for Different Number of Decoder Layers in Self-interpretable Classifier:* In this study, the influence of the decoder architecture on the performance of our self-interpretable classifier was investigated for the case where the black-box classifier was fixed as VGG16. Here, the task of cardiomegaly detection was considered. The primary objective of this study was to explore whether varying the number of layers in the decoder network has a significant impact on the model's classification performance. To achieve this, different decoder architectures were employed with varying number of convolutional layers in the Deconv block. The experiments were performed with 0, 2, 3, 5, and 7 layers in each Deconv block. Table IX shows the performance did not improve after 5 layers for this task.

TABLE IX: Performance (%) with different number of convolutional layers in the Deconv block of the self-interpretable classifier's decoder

| Task: Cardiomegaly Detection | 0 | 2 | 3 | 5 | 7 |
|---|---|---|---|---|---|
| Accuracy | 60.1 | 63.2 | 78.1 | 81.0 | 81.0 |

*3) Examination of False positive (FP) and false negative (FN) Cases:* As depicted in Figure 18, examples are provided from the mammography dataset and the chest X-ray dataset to illustrate FP and FN cases when the VGG16 feature extraction network was used as the encoder of the self-interpretable classifier. Similar to the finding in Sec. IV-G, it was observed that FN cases fail to accurately locate abnormal regions within the images. In contrast, FP cases exhibited E-maps that beared a closer resemblance to E-maps of the true positive class.

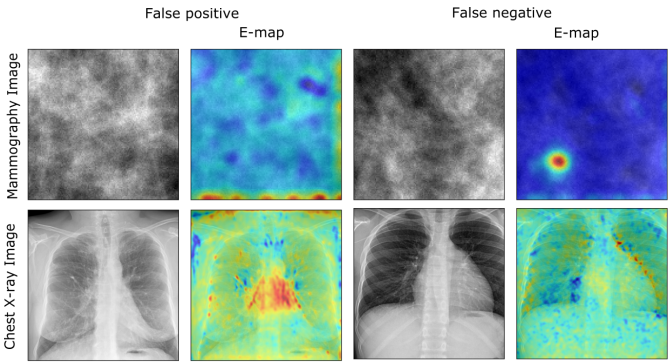

Fig. 18: False positive (FP) and false negative (FN) Analysis. FP cases look similar to Positive class E-map, FN cases look similar to negative class E-map

*4) Stability Analysis of Equivalency Maps:* Similar to the analysis in Sec. IV-H, Fig. 19 shows the stability of the E-maps with different random weight initializations of the self-interpretable network. The E-maps look similar for different random restarts.

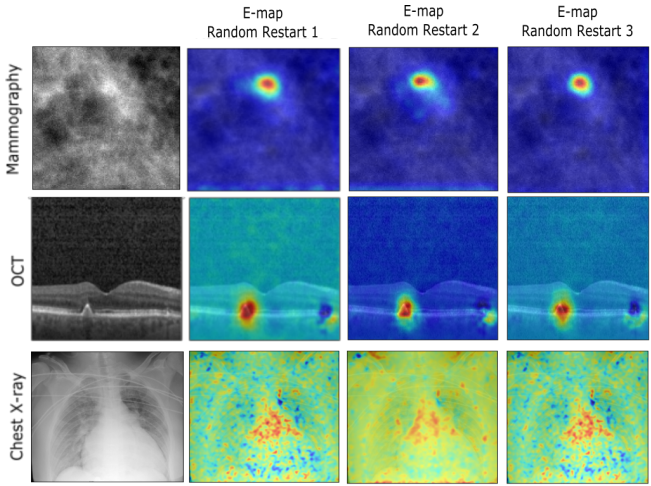

Fig. 19: From the left of each row: input image and E-maps for 3 different random weight initializations of the self-interpretable network. The feature extraction component of the trained VGG16 was used as the encoder. This study demonstrates the stability of the E-maps for different random restarts of the network.

*5) Pixel Intensity Distribution Analysis:* In Fig. 20, the histogram is plotted for positively contributing pixels of the E-maps of normal and abnormal images of different tasks by the self-interpretable network with pre-trained VGG16 encoder. Similar to the findings in Sec. IV-I, it can be seen that there is a significant difference in positively contributing element values for abnormal cases compared to the normal images.

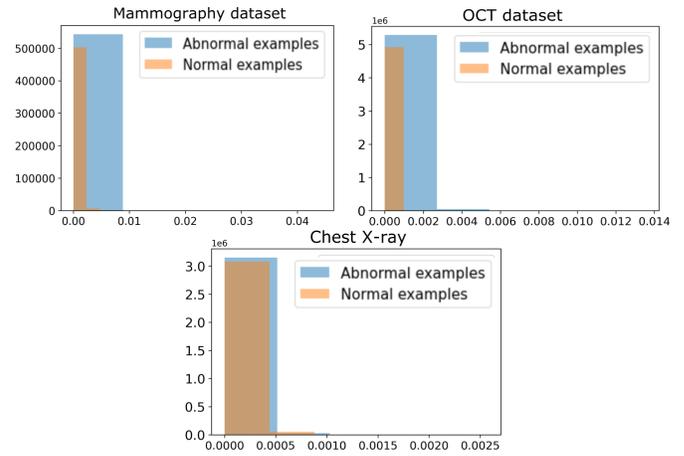

Fig. 20: The histograms of normal vs abnormal cases for each task by the self-interpretable network are shown here. The feature extraction component of the trained VGG16 was used as the encoder. The results show the difference in positively contributing elements for abnormal compared to normal cases.

*6) Quantitatively Evaluating Interpretability: E-map Contribution from Abnormality Regions:* Similar to the baseline CNN results shown in Sec. IV-J, Fig. 21 reveals how the top 1% contributing pixels (red) overlap with the abnormal locations for mammography and chest X-ray dataset respectively. The percentage overlap is written above each image.

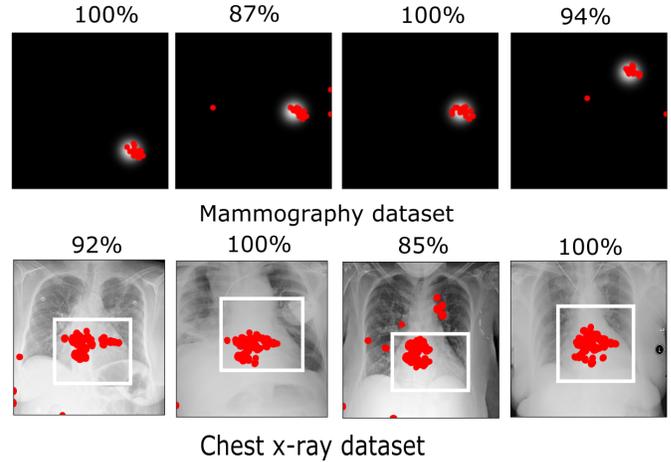

Fig. 21: Examples of overlap between most contributing 1% elements of an E-map and the abnormal region. The feature extraction component of the trained VGG16 was used as the encoder. The top row presents tumor mammography images, and the bottom row presents cardiomegaly chest X-ray images. The red pixels are from the E-map. In most cases, the E-map achieved a high overlap percentage.

### D. Results with ResNet, InceptionV3 and DenseNet

In this study, the outcomes were obtained with the self-interpretable network when the black-box classifiers were ResNet, InceptionV3, and DenseNet. The results of the self-interpretable network corresponding to the ResNet, InceptionV3 and Densenet classifiers are in Table X. It was found that the self-interpretable model achieved very similar accuracy for all three cases. Here, the task of tumor detection using mammography images was considered. Figure 22 shows the abnormal class mammography image and the corresponding E-maps for different networks.

TABLE X: Classification accuracy (%) when different black-box classifiers were considered.

| ResNet | | InceptionV3 | | DenseNet | |
|---|---|---|---|---|---|
| Black-box Classifier | Self-interpretable Classifier | Black-box Classifier | Self-interpretable Classifier | Black-box Classifier | Self-interpretable Classifier |
| 77.9 | 77.8 | 80.1 | 79.8 | 79.95 | 79.5 |

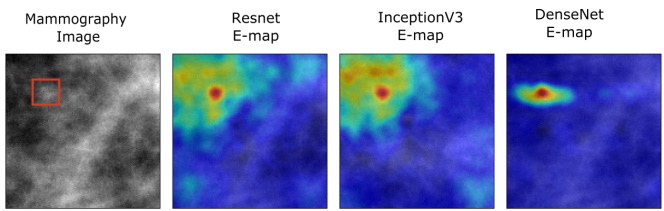

Fig. 22: From left: the tumor class mammography image, E-maps for three different self-interpretable networks corresponding to three different black-box classifiers. the red bounding box shows the location of the tumor.

### E. Existence of Zero-padded Trivial Solution

It is to be noted that to achieve the same performance as the black-box classification network, the decoder of the self-interpretable network can be a simple function that pads zeros to the latent representation yielded by the encoder. In this zero-padding scenario, the E-map would not provide any useful information regarding model interpretability.

To address this concern, we conducted experiments specifically designed to investigate the zero-padding scenario. For these experiments, we selected the task of tumor detection using mammography data. There were two main goals of the studies. The first goal was to demonstrate that the zero-padded solution can indeed be approximated in a contrived setting. The second goal was to demonstrate that the zero-padded solution was not obtained as soon as the contrived settings were removed.

In the first study, we considered an encoder-decoder architecture that closely resembles our interpretable model. The encoder corresponded to a given black-box classifier's feature extraction network. However, the final dense layer responsible for summing the elements of the decoder's output was removed. The purpose of this model was only to predict the zero-padded output of the latent representation of the encoder in a fully supervised way. It should be noted that this experiment is contrived and only seeks to investigate the feasibility of the encoder-decoder model to produce a solution that approximates the zero-padded solution's latent representation. To produce targets for the model training, the latent representations of the input images yielded by the encoder were generated. Next, the latent representations were zero-padded to obtain the target output. Here, the decoder weights were initialized randomly. Finally, the model was trained in a supervised manner to predict that zero-padded latent embedding from the corresponding individual input image. We found that the model was able to accurately estimate the zero-padded solution.

Next, we revisited the complete self-interpretable encoder-decoder model, including the final dense layer, and trained the model to predict the test statistic value of the black box classifier as described in the manuscript. Here, the decoder weights were initialized using the decoder weights previously trained in the contrived experiment above. It was observed that under these conditions, the decoder's output or E-map closely resembled a zero-padded version of the latent representation from the encoder.

Subsequently, we perturbed the decoder weights by adding Gaussian noise of 0 mean and varying standard deviation levels of 0.01, 0.03, 0.07, 0.09, 0.1, 0.2, and 0.3. In this case, we observed that the network did not produce a zero-padded solution after a standard deviation level of 0.1.

Together, these experiments corroborated our claim that our method is unlikely to produce the zero-padded solution, in general, except in specific cases of decoder weight initialization. Our empirical investigation suggests that such an extremely specific weight initialization is unlikely to occur during the training of our interpretable model when the pre-training is based on commonly used random initialization methods, such as the uniform or normal distribution.

### ACKNOWLEDGEMENT

The authors would like to thank Dr. FJ. Brooks for coining the term "Equivalency Map".

### REFERENCES

[1] A. Singh, S. Sengupta, and V. Lakshminarayanan, "Explainable deep learning models in medical image analysis," *Journal of Imaging*, vol. 6, no. 6, p. 52, 2020.

[2] Y. Zhang, P. Tiňo, A. Leonardis, and K. Tang, "A survey on neural network interpretability," *IEEE Transactions on Emerging Topics in Computational Intelligence*, 2021.

[3] C. Rudin, "Stop explaining black box machine learning models for high stakes decisions and use interpretable models instead," *Nature Machine Intelligence*, vol. 1, no. 5, pp. 206–215, 2019.

[4] Y. Wang and X. Wang, "Self-interpretable model with transformation equivariant interpretation," *Advances in Neural Information Processing Systems*, vol. 34, pp. 2359–2372, 2021.

[5] M. Du, N. Liu, and X. Hu, "Techniques for interpretable machine learning," *Communications of the ACM*, vol. 63, no. 1, pp. 68–77, 2019.

[6] K. Simonyan, A. Vedaldi, and A. Zisserman, "Deep inside convolutional networks: Visualising image classification models and saliency maps," *arXiv preprint arXiv:1312.6034*, 2013.

[7] J. T. Springenberg, A. Dosovitskiy, T. Brox, and M. Riedmiller, "Striving for simplicity: The all convolutional net," *arXiv preprint arXiv:1412.6806*, 2014.

[8] R. R. Selvaraju, M. Cogswell, A. Das, R. Vedantam, D. Parikh, and D. Batra, "Grad-cam: Visual explanations from deep networks via gradient-based localization," in *Proceedings of the IEEE international conference on computer vision*, pp. 618–626, 2017.

[9] M. Sundararajan, A. Taly, and Q. Yan, "Axiomatic attribution for deep networks," in *International conference on machine learning*, pp. 3319–3328, PMLR, 2017.

[10] M. T. Ribeiro, S. Singh, and C. Guestrin, ""why should i trust you?" explaining the predictions of any classifier," in *Proceedings of the 22nd ACM SIGKDD international conference on knowledge discovery and data mining*, pp. 1135–1144, 2016.

[11] A. Binder, G. Montavon, S. Lapuschkin, K.-R. Müller, and W. Samek, "Layer-wise relevance propagation for neural networks with local renormalization layers," in *Artificial Neural Networks and Machine Learning–ICANN 2016: 25th International Conference on Artificial Neural Networks, Barcelona, Spain, September 6-9, 2016, Proceedings, Part II 25*, pp. 63–71, Springer, 2016.

[12] S. Jain, S. Wiegreffe, Y. Pinter, and B. C. Wallace, "Learning to faithfully rationalize by construction," *arXiv preprint arXiv:2005.00115*, 2020.

[13] D. Alvarez Melis and T. Jaakkola, "Towards robust interpretability with self-explaining neural networks," *Advances in neural information processing systems*, vol. 31, 2018.

[14] P. W. Koh, T. Nguyen, Y. S. Tang, S. Mussmann, E. Pierson, B. Kim, and P. Liang, "Concept bottleneck models," in *International Conference on Machine Learning*, pp. 5338–5348, PMLR, 2020.

[15] C. Chen, O. Li, D. Tao, A. Barnett, C. Rudin, and J. K. Su, "This looks like that: deep learning for interpretable image recognition," *Advances in neural information processing systems*, vol. 32, 2019.

[16] R. Agarwal, L. Melnick, N. Frosst, X. Zhang, B. Lengerich, R. Caruana, and G. E. Hinton, "Neural additive models: Interpretable machine learning with neural nets," *Advances in Neural Information Processing Systems*, vol. 34, pp. 4699–4711, 2021.

[17] S. Mohammadjafari, M. Cevik, M. Thanabalasingam, and A. Basar, "Using protopnet for interpretable alzheimer's disease classification.," in *Canadian Conference on AI*, 2021.

[18] B. Chidester, T. Zhou, M. N. Do, and J. Ma, "Rotation equivariant and invariant neural networks for microscopy image analysis," *Bioinformatics*, vol. 35, no. 14, pp. i530–i537, 2019.

[19] D. S. Kermany, M. Goldbaum, W. Cai, C. C. Valentim, H. Liang, S. L. Baxter, A. McKeown, G. Yang, X. Wu, F. Yan, *et al.*, "Identifying medical diagnoses and treatable diseases by image-based deep learning," *Cell*, vol. 172, no. 5, pp. 1122–1131, 2018.

[20] C. Castella, K. Kinkel, F. Descombes, M. P. Eckstein, P.-E. Sottas, F. R. Verdun, and F. O. Bochud, "Mammographic texture synthesis: second-generation clustered lumpy backgrounds using a genetic algorithm," *Optics express*, vol. 16, no. 11, pp. 7595–7607, 2008.

[21] M. Ruschin, A. Tingberg, M. Båth, A. Grahn, M. Håkansson, B. Hemdal, and I. Andersson, "Using simple mathematical functions to simulate pathological structures—input for digital mammography clinical trial," *Radiation protection dosimetry*, vol. 114, no. 1-3, pp. 424–431, 2005.

[22] X. Wang, Y. Peng, L. Lu, Z. Lu, M. Bagheri, and R. M. Summers, "Chestx-ray8: Hospital-scale chest x-ray database and benchmarks on weakly-supervised classification and localization of common thorax diseases," in *Proceedings of the IEEE conference on computer vision and pattern recognition*, pp. 2097–2106, 2017.

[23] A. Krizhevsky, I. Sutskever, and G. E. Hinton, "Imagenet classification with deep convolutional neural networks," *Communications of the ACM*, vol. 60, no. 6, pp. 84–90, 2017.

[24] D. P. Kingma and J. Ba, "Adam: A method for stochastic optimization," *arXiv preprint arXiv:1412.6980*, 2014.

[25] D. Smilkov, N. Thorat, B. Kim, F. Viégas, and M. Wattenberg, "Smoothgrad: removing noise by adding noise," *arXiv preprint arXiv:1706.03825*, 2017.

[26] X. Glorot and Y. Bengio, "Understanding the difficulty of training deep feedforward neural networks," in *Proceedings of the thirteenth international conference on artificial intelligence and statistics*, pp. 249–256, JMLR Workshop and Conference Proceedings, 2010.

[27] Z. Wang, A. C. Bovik, H. R. Sheikh, and E. P. Simoncelli, "Image quality assessment: from error visibility to structural similarity," *IEEE transactions on image processing*, vol. 13, no. 4, pp. 600–612, 2004.

[28] O. Li, H. Liu, C. Chen, and C. Rudin, "Deep learning for case-based reasoning through prototypes: A neural network that explains its predictions," in *Proceedings of the AAAI Conference on Artificial Intelligence*, vol. 32, 2018.